# Surface doping in T6/ PDI-8CN$_2$ Heterostructures investigated by transport and photoemission measurements

L. Aversa[1], R. Verucchi[1], R. Tatti[1], F. V. Di Girolamo[2,a], M. Barra[2], F. Ciccullo[2], A. Cassinese[2] and S. Iannotta[1,3]

[1] Istituto dei Materiali per l'Elettronica ed il Magnetismo, IMEM-CNR, sezione FBK di Trento, Via alla Cascata 56/C - Povo, 38123 Trento, Italy

[2] CNR-SPIN and Department of Physics Science, University of Naples "Federico II", P.le Tecchio 80, 80125 Naples, Italy

[3] Istituto dei Materiali per l'Elettronica ed il Magnetismo, IMEM-CNR, Parco Area delle Scienze 37/A - 43124 Parma, Italy

[a] email: fdigirolamo@na.infn.it

**ABSTRACT.**

In this paper, we discuss the surface doping in sexithiophene (T6) organic field-effect transistors by N,N'-bis (n-octyl)-dicyanoperylenediimide (PDI-8CN$_2$). We show that an accumulation heterojunction is formed at the interface between the organic semiconductors and that the consequent band bending in T6 caused by PDI-8CN$_2$ deposition can be addressed as the cause of the surface doping in T6 transistors. Several evidences of this phenomenon have been furnished both by electrical transport and photoemission measurements, namely the increase in the conductivity, the shift of the threshold voltage and the shift of the T6 HOMO peak towards higher binding energies.

The creation and control of electronic properties of organic/organic interfaces is receiving great attention for several applications in electronics and optoelectronics. The electronic properties of these systems are ruled by the alignment of the energy levels of the organic species, showing narrower and spatially more localized bands with respect to inorganics. In particular, the Fermi energy difference between the dopant and the semiconductor can drive a charge transfer process which leads to the enrichment of the majority charge carriers in the semiconductor. This occurrence enhances the conductivity of the overall system and consequently represents an efficient way to dope organic semiconductors, improving their electrical properties [1-2].

The presence of charge transfer and accumulation layers is currently under investigation for several pairs of organic conjugated materials [3,4,5]. To this aim, useful hints can be given by electrical measurements [6,7,8]. However, the most viable approach to verify and quantify their occurrence is the Ultraviolet Photoemission Spectroscopy (UPS) [2][9]. This technique, indeed, enables the acquisition of most of the critical parameters, such as the Fermi level position, the vacuum and bands energy positions and consequently work functions and expected position of empty states. Hence, the presence of interface dipoles and charge transfer processes can be properly defined by changes in levels position and formation of band bending.

In a previous report, it was shown that the formation of an accumulation junction is able to explain the physical and electrical properties exhibited by T6/PDI-8CN$_2$ heterostructures [10]. These results resemble those discussed in other recent reports [1,2,3,6,7,8], with the most remarkable example given by the insulator-to-metal transition demonstrated at the interface between Tetrathiofulvalene (TTF) and 7,7,8,8-tetracyanoquinodimethane (TCNQ) single crystals [11][12].

In this paper, we have analyzed the surface doping effect produced by the evaporation of few nanometers of PDI-8CN$_2$ on the top of 15 nm thick sexithiophene (T6) films, acting as active channels of field-effect transistors. Both electrical transport and photoemission measurements were performed. The electrical characterization evidenced the large enhancement of the channel conductance and the shift in of threshold voltages as extracted from the transfer-curves in the PDI-8CN$_2$ doped T6 transistors. These findings suggest the occurrence of band-bending at the interface between T6 and PDI-8CN$_2$, as confirmed by the Ultraviolet Photoelectron Spectroscopy (UPS) investigation, that also put in evidence the presence of interface dipoles.

Indeed, UPS measurements assess the upward shift of the Highest Occupied Molecular (HOMO) Level in T6 and the downward shift of the calculated Lowest Unoccupied Molecular Orbital (LUMO) in PDI-8CN$_2$, demonstrating clearly the formation of an accumulation junction at the T6/PDI-8CN$_2$ heterointerface.

The organic field-effect transistors were fabricated by in-situ sequential deposition of each organic layer on Si$^{++}$(500 µm)/SiO$_2$(200 nm) substrates, provided of gold Source-Drain interdigitated electrodes (bottom-contact bottom-gate configuration). The thickness of the deposited films was measured by a thickness monitor. More details about the transistor layout can be found elsewhere [10].

Before the deposition of the organic layers, the substrates were cleaned in ultrasonic baths of acetone and ethanol, followed by drying in pure N$_2$ gas. The organic films were deposited in a high vacuum evaporation system equipped with independent Knudsen-cells filled with T6 (Sigma Aldrich) and PDI-8CN$_2$ (ActivInk 1200 Polyera Corporation Inc.) powders. A thick (15 nm) layer of T6 was firstly deposited, while, afterwards, thin (1 nm and 4 nm) PDI-8CN$_2$ films were evaporated. As reference, single-layer samples with 15 nm of T6, 4 nm of PDI-8CN$_2$ and 20 nm of PDI-8CN$_2$ (bulk) were grown directly on the Au/SiO$_2$ substrate. The film thickness was monitored by a quartz microbalance. The samples were analysed without any ex-situ chemical treatment or in-situ cleaning procedure (i.e. sputtering) in order to avoid possible changes of their electronic properties. The electrical characterizations were performed in vacuum (10$^{-5}$ mbar) and in dark conditions using a Janis probe station. I$_{DS}$ vs V$_{DS}$ (drain-source current vs drain-source voltage at fixed gate-source voltages) output curves and I$_{DS}$ vs V$_{GS}$ (drain-source current vs gate-source voltage at fixed drain-source voltages) transfer-curves were recorded using a Keithley 2612A Dual-Channel system source-meter instrument.

The surface morphology of the samples was analyzed by a XE100 Park AFM microscope operating in air. Images were acquired using doped silicon cantilevers (resonance frequency around 300 KHz) provided by Nanosensor$^{TM}$, in non-contact mode. The root-mean-square roughness ($R_q$) of the film surface was determined by the Park XEI Software as the standard deviation of the film height distribution.

UPS analysis was performed with a He UV lamp (HeI, hν = 21.2 eV) and a PSP electron energy analyser, leading to a total energy resolution <100 meV. The valence band (VB) binding energy (BE) was referred to the Au Fermi level. The two bilayer configurations were analyzed, together with reference samples of T6, PDI-8CN$_2$ (20 nm and 4 nm thick film) and Au (sputtered thick film). The metal and organic work function (WF) was evaluated from the position of the secondary electron cut-off (SECO), while the organic ionization potential (IP) was evaluated (±0.1 eV uncertainty) from the difference between the photon energy and the spectrum length. The analysis system enabled photoemission from a film spot larger compared to the interdigitated electrodes, so that VB are representative of organics covering both metal contacts and oxide region.

In Fig. 1, the AFM images of four samples (a: 15 nm thick T6 single layer; b: T6(15nm)/PDI-8CN$_2$ (1nm); c: T6(15nm)/PDI-8CN$_2$ (4nm); d: 4 nm thick PDI-8CN$_2$ single layer) are reported. As shown in Fig.1a, the T6 thick layer exhibits mainly the classical three-dimensional island morphology based on the coalescence of disk-shaped crystallites [13]. The film surface is terraced with a molecular step of about 2.4 nm, while the root-mean-square roughness ($R_q$), estimated by XEI software (Park systems), is slightly higher than 2 nm ($R_q \approx 2.1$ nm).

The surface of the thin (4nm) PDI-8CN$_2$ film (fig.1d) results very smooth with $R_q$ of only 0.98 nm. Indeed, this film is approximately given by the overlapping of two complete molecular monolayers, as a consequence of the PDI-8CN$_2$ capability to follow a layer-by-layer growth mode when deposited in form of very thin layers on SiO$_2$ [14, 15].

On the other hand, the morphology of PDI-8CN$_2$ thin films is considerably modified when they are evaporated on the 15 nm thick T6 bottom layer. In this situation, due to the T6 island-like structure, PDI-8CN$_2$ appears to be unable to uniformly cover the organic substrate beneath and the surface roughness is found to noticeably increase in comparison with T6 single layer. This is especially true for the (PDI-8CN$_2$ 1nm)/( T6 15nm) heterostructure which surface roughness $R_q$ is 4.14 nm. This experimental data suggests also that, in this case, PDI-8CN$_2$ layer is not continuous and does not cover completely the T6 layer, as also demonstrated by the electrical characterizations (see below). Finally, for the (PDI-8CN$_2$ 4nm)/( T6 15nm), $R_q$ is reduced down to 2.64 nm, indicating that PDI-8CN$_2$ tends to mainly fill the spaces among the T6 islands.

Fig.2a shows the output curves measured for the single layer (15 nm thick ) T6 transistor and for the (PDI-8CN$_2$ 1nm)/(T6 15nm) and (PDI-8CN$_2$ 4nm)/(T6 15nm) layered devices. These measurements were recorded by applying $V_{GS}$=0 V and sweeping $V_{DS}$ from 0 to 50 V. From these data, it is possible to observe a monotonic increase of the $I_{DS}$ current flowing in the device with the thickness of the PDI-8CN$_2$ layer. This experimental evidence is a demonstration of the surface doping effect generated in the T6 layer by the presence of PDI-8CN$_2$ molecules [1, 3, 16]. Indeed, these $I_{DS}$ curves are related only to the flow in the T6 film of holes injected form the drain electrode, while any electron $I_{DS}$ current in the PDI-8CN$_2$ layers should follow a saturating behaviour at high $V_{DS}$ getting values not exceeding few µAs (see Fig.3 in the reference[10]).

This conclusion is further supported by the increase with the thickness of the $I_{DS}$ current illustrated in the inset of Fig. 2, and extracted for $V_{GS}$ = 0 V and $V_{DS}$ comprised between -1 V and 1 V. In these conditions, indeed, the channel resistance should not be sensitively affected by the field-effect phenomenon and consequently correspond to the bulk value [1].

The transfer-curves, reported in Fig. 2b, are recorded in saturation region ($V_{DS}$= -50 V and $V_{GS}$ sweeping from 0 to -50 V). The behaviour is typical of p-type devices, as expected since the thickness of T6 is higher than PDI-8CN$_2$. Nevertheless, in the curve recorded for the device with 4 nm of PDI-8CN$_2$ it is possible to observe a slight increase of the current with positive voltages, which could be an indication of n-type behaviour. No evidences of ambipolar behaviour are instead observable in the curve recorded for the device realized with the deposition of 1 nm of PDI-8CN$_2$, as also the non uniformity of the layer evidenced by the AFM images suggests. This observation is remarkable, since the presence of two conductive channels as an explanation for the increased conductivity can be excluded, at least in this case.

The threshold voltage also increases monotonously with the thickness of the PDI-8CN$_2$ layer, which represents a further indication of a surface transfer doping effect [1, 3, 16].

The mobility of the T6, extracted with the traditional MOSFET model, increases from $5 \times 10^{-3}$ to $2 \times 10^{-2}$ cm$^2$/volt*sec with the addition of 1-4 nm of PDI-8CN$_2$. It is worth to mention that the mobility values are only effective ones, and are used for the sake of comparison.

VB spectra are shown in Fig. 3b, with close-up of the HOMO (Fig. 3c) and SECO (Fig. 3a) energy regions. Common features for all organic films are the absence of any Au VB related signal and the significant shift of the SECO position with respect to the metal surface, suggesting the presence of a dipole (Δ) at the organic/Au interface [4, 17]. PDI-8CN$_2$ VB is characterized by several structures in the 4-12 eV BE range, with HOMO located at ~2.9 eV, IP of 7.1 eV and WF = 4.8 eV. Other studies for unsubstituted perylene molecule report a HOMO located at 1.8eV and an IP of 5.1 eV [18], while theoretical analysis led to an IP = 6.7 eV [19]. According to what observed for CN-substituted phthalocyanine [20], the downward shift of the HOMO peak and the IP increase with respect to the unsubstituted molecule can be related to the presence of the cyano functional groups. Comparing VBs of the 4 nm and thick perylene films (Fig. 3), a +0.3 eV BE shift and a broad underlying band in the 4-12 eV energy region are present, IP is the same but WF is lower (-0.3 eV). T6 VB shows HOMO and HOMO-1 located at 1.7 eV and 2.4 eV, and is characterized by a very low intensity. This is typical of an island growth in organic films, with high aspect ratio as indeed observed by AFM analysis. The HOMO BE is in agreement with previous studies on Au substrate [21], while similar WFs have been reported for T6 films grown on SiO$_2$ [22], but IP is significantly lower [23, 24, 25]. Similar results have been attributed to specific molecular packing, leading to greater π conjugation and polarizability and thus to the observed variation of IP and WF [22].

In bilayer films (Fig. 3), the presence of PDI-8CN$_2$ significantly changes the T6 VB as features of both species can be identified in the spectra. The main characteristics of the two VBs are the intensity enhancement of the HOMO T6 and perylene features (with respect to the single reference thick layers), respectively located at 1.7-1.3 eV and 3.5-3.1 eV for the 1-4 nm PDI-8CN$_2$/T6 films (Fig. 3c). The 4 nm PDI-8CN$_2$/T6 VB well resembles the structures of a +0.2 eV BE shifted 4 nm perylene film. At the higher perylene film thickness, the intensity of T6 features decreases, suggesting a progressive covering of the T6 islands only for this bilayer. The shift of the SECO position (see Fig. 3a) reveals IP and WF increasing with perylene thickness.

Combining WF, IP, Δ, HOMO centroid and leading edge energy, expected LUMO position, we realized a scheme reproducing the vacuum level position and band alignment for all analyzed films and Au (see Fig. 4) [3,4,17]. The accepted HOMO-LUMO energy gap for PDI-8CN$_2$ is ~2.4 eV [19], while for T6 is ~2.1 eV [26]. The perylene/Au films are characterized by a LUMO position really close to the metal Fermi level, leading to a very low electron injection barrier from Au. Comparing the 4 nm and thick PDI-8CN$_2$ films, the -0.3 eV BE shift of the HOMO suggests presence of band bending, but the most important evidence is the presence of a peak in the HOMO-LUMO gap (see arrow in Fig. 3c), probably a filled LUMO' or relaxed HOMO' peak that is definitely due to a charge transfer from the inorganic surfaces and accumulation in the organic film [3,17]. The charge transfer from the SiO$_2$ surface is unlikely, as already reported [27]. Thus, the electrons injection concerns the organic/Au interface and is the origin of the whole WF +0.3 eV increase when the thermodynamic equilibrium is reached in the thick PDI-8CN$_2$ film, where the LUMO'/HOMO' feature disappears (confirming its origin related to a charge transfer at the interface) and LUMO is aligned with Fermi level. We can argue the thickness $w$ of the organic layer interested by the charge accumulation to be at least 4 nm. The transferred charge density $N_D$ in a classical approach valid for inorganics is equal to $2\varepsilon_0\varepsilon_R\phi/ew^2$, where $\varepsilon_0$ and $\varepsilon_R$ are the vacuum and relative permittivity, $\phi$ is the observed band bending. Considering the equivalence of relative permittivity and dielectric constant in static conditions, $\varepsilon_R = 2.73$ can be used for perylene [28]. $N_D$ has a value $<= 5.7 \times 10^{18}$ corresponding to $\sim <= (7 \times 10^{-3})e$ per molecule for a molecular density of about $8 \times 10^{20}$ mol·cm$^{-3}$ [28]. Comparing T6 thick film energy levels with previous studies of growth on Au [21] and SiO$_2$ [22], our results are in agreement with a reduction of organic IP and inorganic WF, as well as with formation of interface dipoles and absence of charge transfer [23, 24, 25]. Nevertheless, HOMO BE and WF are more similar to those obtained for the T6/Au case [21], suggesting that the metal/organic interface plays a major role in defining the final T6 film electronic properties. The deposition of 1 nm PDI-8CN$_2$ layer on the thick sexithiophene film leads to a +0.2 eV increase of both IP and WF, but no T6 HOMO (and consequently LUMO) shift, according to absence of band bending and presence of a $\Delta\phi = -0.2$ eV surface dipole. Perylene HOMO BE is +0.5 eV than in the reference film, with the corresponding LUMO lying 0.7 eV above T6 HOMO leading edge. In the 4 nm PDI-8CN$_2$/T6 film, a $\Delta = -1.3$ eV leads to an IP similar to T6 and a WF similar to perylene. T6 and PDI-8CN$_2$ HOMOs show a shift towards lower BE of -0.5 eV, with a perylene LUMO located 0.1 eV below the Fermi level. The BE shifts suggest the presence of both band bending and interface dipole, the latter resulting to be $\Delta\phi = -0.8$ eV, considering the observed WF increase. The charge transfer involves the PDI-8CN$_2$ LUMO that could receive electrons from the T6 HOMO or the metal. The absence of any Au VB signal in the T6/Au film suggests presence of a 4-5 nm layer of T6 over the whole metal surface, also among islands, thus avoiding any significant PDI-8CN$_2$/Au interaction also through tunnelling processes [4]. On the other hand, the T6 HOMO BE shift is an evidence of the organic involvement in the charge transfer process [3,4,17,21] and is a further indication that the increase in the source drain current in the output and transfer curves is completely due to a surface doping of T6 by PDI-8CN$_2$. Comparing results for the two analyzed films with 4 nm perylene layer thickness, the HOMO position in the bilayer is closer to that of a thick film in thermal equilibrium, suggesting that charge transfer occurred more efficiently than in the corresponding 4 nm PDI-8CN$_2$/Au case, where LUMO is still 0.3 eV below the Fermi level. Due to the superposition of several molecular orbitals in the VB, it is not possible to find evidence of the presence of the LUMO'/HOMO' feature as was in the organic/metal interface. The effects of surface morphology must be taken into account, with the first 1 nm of perylene probably covering zones between T6 islands (with no significant change in the electronic properties) and the subsequent 3 nm achieving a complete T6 coverage (inducing all observed phenomena and observed charge transfer). On the basis of the proposed classical metal/semiconductor interface, for PDI-8CN$_2$ the charge (electron) density in the accumulation layer $N_D$ corresponds to $1.7 \times 10^{19}$ cm$^{-3}$, considering $w = 30$ Å, $\phi = 0.5$ eV. Given the total charge conservation, the calculated $N_D$ charge (hole) density for T6 leads to an estimated accumulation layer thickness $w = 57$ Å, considering $\varepsilon_R = 10$ [22]. The charge per molecule for T6 and perylene are $\sim(3.5 \times 10^{-2})e$ and $\sim(2 \times 10^{-2})e$, considering a molecular density of $4.8 \times 10^{20}$ and $8 \times 10^{20}$ mol·cm$^{-3}$ respectively [22, 28]. The proposed level alignment at the heterojunction is in agreement with other studied organic systems [2,29-31] and confirms results for the PDI-8CN$_2$/T6 bilayers (with different thicknesses) achieved by some of the authors of this study, where a substantial charge transfer process was supposed between the two organics [10].

In this paper, the formation of an accumulation heterojunction at the PDI-8CN$_2$ and T6 has been demonstrated to represent an efficient tool for the surface doping in field-effect transistors. It was investigated by transport and ultraviolet photoemission spectroscopy (UPS) measurements. The transport measurements give several indications of the formation of an accumulation heterojunction. First of all, from the output curves at zero gate a monotonous decrease of the resistance at increasing PDI-8CN$_2$ thicknesses is observed, which is not simply caused by the contemporary presence of both charge carriers in the channel and is consequently due to an enrichment of the hole density at the heterointerface. Moreover, the threshold voltages extracted from the transfer curves also suggest that the accumulated holes are filling the traps at the heterointerface [8]. UPS analysis reveals presence of a significant charge transfer from the T6 HOMO to the perylene LUMO, more efficient than was in the case of the PDI-8CN$_2$/Au interface.

In the T6 an accumulation layer of 5-6 nm can be estimated by considering conventional semiconductors interface models.

The findings, together with the previous works published on this subject [10] suggest that, exploiting the charge transfer at the heterointerface, innovative organic devices based on accumulation heterojunctions can be developed.

.

Financial support from the Italian Ministry of Research under PRIN 2008 project 2008FSBKKL "*Investigation of n-type organic materials and related devices of interest for electronic applications"* is gratefully acknowledged.

**References**

[1] Y. Abe, T. Hasegawa, Y. Takahashi, T. Yamada, and Y. Tokura, Appl. Phys. Lett. **87** 153507 (2005).

[2] Shenghao Wang, Takeaki Sakurai, Ryusuke Kuroda, and Katsuhiro Akimoto Appl. Phys. Lett. 100, 243301 (2012)

[3] Wei Chen, Dongchen Qi, Xingyu Gao, Andrew Thye Shen Wee Progress in Surface Science 84, 279 (2009).

[4] S.Braun, W. R. Salaneck, M. Fahlman, Adv. Mat. 21, 1450 (2009).

[5] D. Çakir, M. Bokdam, M. P. de Jong, M. Fahlman, G. Brocks, Appl. Phys. Lett. 100, 203302 (2012)

[6] F. Zhu, K. Lou, L. Huang, J. Yang, J. Zhang, H. Wang, Y. Geng, and D. Yan, Appl. Phys. Lett. **95**, 203106 (2009).

[7] F. Zhu, J. Yang, D. Song, C. Li, and D. Yan. Appl. Phys. Lett. **94**, 143305 (2009).

[8] H. Wang, J. Wang, H. Huang, X. Yan, D. Yan, Organic Electronics **7**, 369 (2006).

[9] K. M. Lau, J. X. Tang, H. Y. Sun, C. S. Lee, S. T. Lee and Donghang Yan, Appl. Phys. Lett **88**, 173513 (2006).

[10] F.V. Di Girolamo, M. Barra, F. Chiarella, S. Lettieri, M. Salluzzo, A. Cassinese, Phys. Rev. B **85**, 125310 (2012).

[11] H. Alves, A. S. Molinari, H. Xie & A. F. Morpurgo, Nature Materials 7, 574 (2008).

[12] M. Nakano, H. Alves, A. S. Molinari, S. Ono, N. Minder,and A. F. Morpurgo Appl. Phys. Lett. **96**, 232102 (2010).

[13] F. Dinelli, M. Murgia, P. Levy, M. Cavallini, F. Biscarini, D.M. DeLeeuw, Phys. Rev. Lett.**92**, 116802 (2004).

[14] F. Chiarella, M. Barra, A. Cassinese , F.V. Di Girolamo, P. Maddalena, L. Santamaria, S. Lettieri, Appl Phys A 104, 39 (2011).

[15] F. Liscio, S. Milita,C. Albonetti, P. D'Angelo, A. Guagliardi, N. Masciocchi, R. G. Della Valle, E. Venuti, A. Brillante, and F. Biscarini Adv. Funct. Mater. 22, 943 (2012).

[16] H. Kleemann , C. Schuenemann, A. A. Zakhidov, M.Riede, B. Lüssem, K. Leo Organic Electronics 13, 58 (2012).

[17] A. Kahn, N. Koch, W. Gao, J. Pol. Sci. B 41, 2529 (2003)

[18] S.J. Kang, Y. Yi, K. Cho, K. Jeong, K.-H. Yoo, C.N. Whang, Synthetic Metal 151, 120 (2005).

[19] B. A. Jones, A. Facchetti, M. R. Wasielewski, T. J. Marks, J. Am. Chem. Soc., 129, 15259 (2007)

[20] K. M. Kadish, K. M. Smith, R. Guilard, *The Porphyrin Handbook Volumes11-20: Vol.16 Phthalocyanines:Spectroscopy and Electrochemical characterization* (Academic Press, New York, 2003), p. 258

[21] Y. Ge, J. E. Whitten, Chem. Phys. Lett. 448, 65 (2007).

[22] J. Ivanco, J. R. Krenn, M. G. Ramsey, F. P. Netzer, T. Haber, R. Resel, A. Haase, B. Stadlober, and G. Jakopic, J. App. Phys. 96, 2716 (2004).

[23] M. Grobosch, M. Knupfer, Organic Electronics 8, 625 (2007).

[24] T. Wagner, D. R. Fritz, P. Zeppenfeld Organic Electronics 12 442 (2011).

[25] A. Chandekar, J.E. Whitten, Synthetic Metals 150, 259 (2005).

[26] O. Pellegrino, M. Rei Vilar, G. Horowitz, F. Kouki, F. Garnier, J.D. Lopes da Silva, A.M. Botelho do Rego, Thin Solid Films 252, 327 (1998).

[27] D. Jin, W. Wanga, A. Rahmana, J. Lizhen, H. Zhang, H. Li, P. Hea, S. Bao, Applied Surface Science 257, 4994 (2011).

[28] M. M. El-Nahhas, H. Abdel-Khalek, E. Salem, Adv. Cond. Matt. Phys., doi:10.1155/2012/698934 (2012).

[29] H. Hintz, H. Peisert, U. Ayg, F. Latteyer, I. Biswas, P. Nagel, M. Merz, S. Schuppler, D. Breusov, S. Allard, U. Scherf, T. Chass, Chem. Phys. Chem., 11, 269 (2010).

[30] J. Frisch, M. Schubert, E. Preis, J. P. Rabe, D. Neher, U. Scherfc, N. Koch, J. Mater. Chem., 22, 4418 (2012).

[31] W. Wang, D. Placencia, N. R. Armstrong, Org. Electr. 12, 383 (2011).

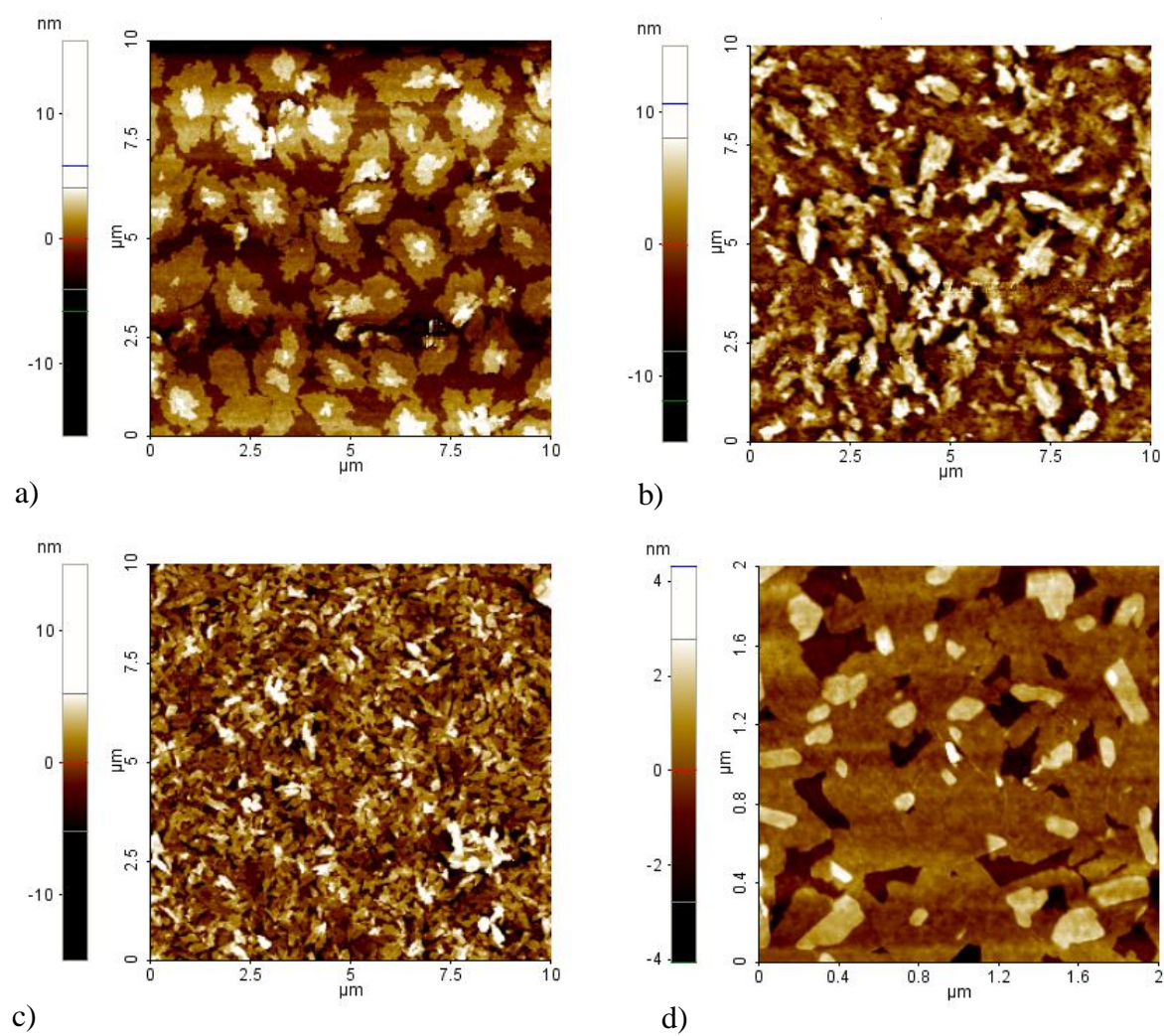

**Fig. 1.** (colour online): 10 x 10 $\mu m^2$ non contact AFM images of (a) 15 nm of T6, ($R_q$ = 2.1 nm) (b) 1 nm and (c) 4 nm of PDI-8CN$_2$ films deposited on 15 nm of T6 ($R_q$ respectively 4.14 nm and 2.64 nm) and (d) 4 nm of PDI-8CN$_2$ ($R_q$ = 0.98 nm).

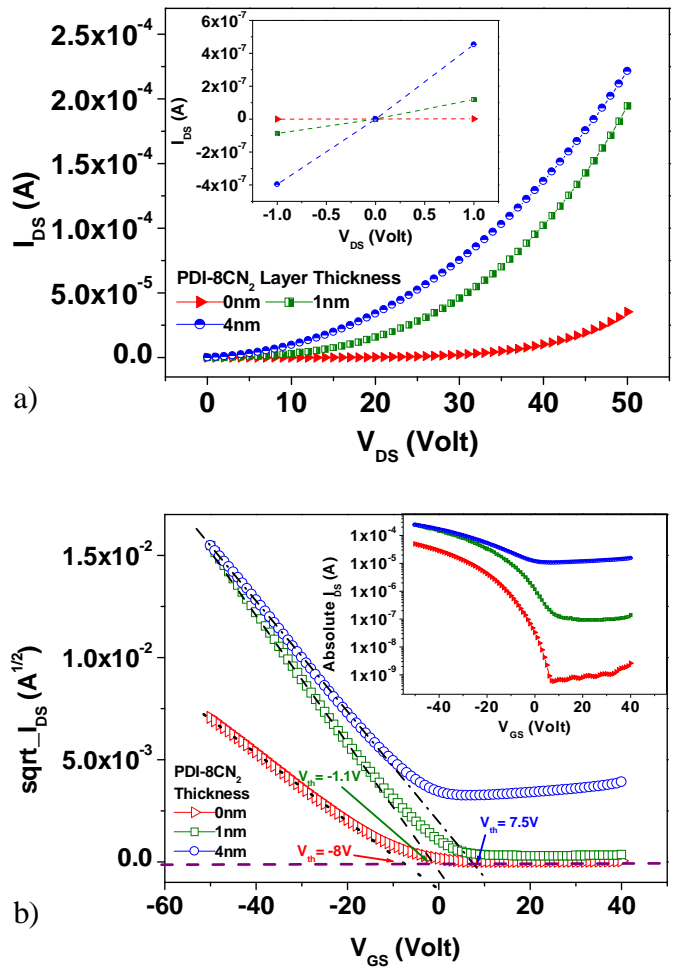

**FIG. 2.** (colour online) Electrical responses measured in vacuum for T6(15nm)/PDI-8CN$_2$ heterostructure field-effect devices with different thickness (0, 1 and 4 nm) of the PDI-8CN$_2$ layer: a) Output-curves ($V_{GS}$ = 0 V and $V_{DS}$ from 0 to 50 V). In the inset, the same curves are magnified for $V_{DS}$ ranging between -1 and 1V; b) Square-root currents measured for the transfer-curves in the saturation regime ($V_{DS}$=50V). In the inset, the transfer-curves in the saturation regime are presented in a semi-log plot.

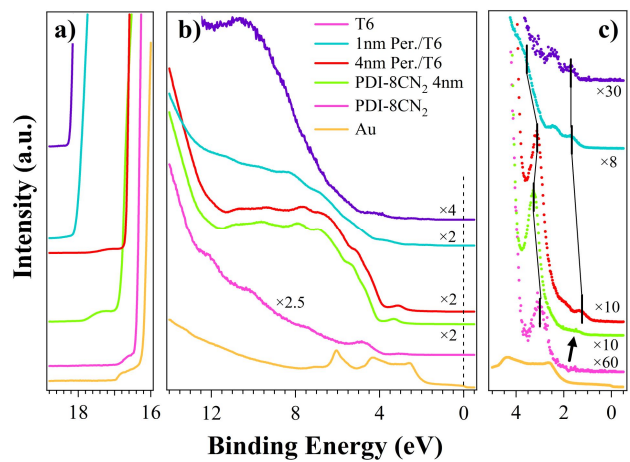

**FIG. 3** Valence band analysis of T6/PDI-8CN$_2$ heterojunctions, in comparison with thick film of T6, perylene (also shown for a 4nm thinner film) and Au. Wide range spectra are shown (b), together with the SECO (a) and the 0-5 eV (c) range, with vertical markers showing HOMO position of T6 and PDI-8CN$_2$ in different films. The arrow shows the gap state.

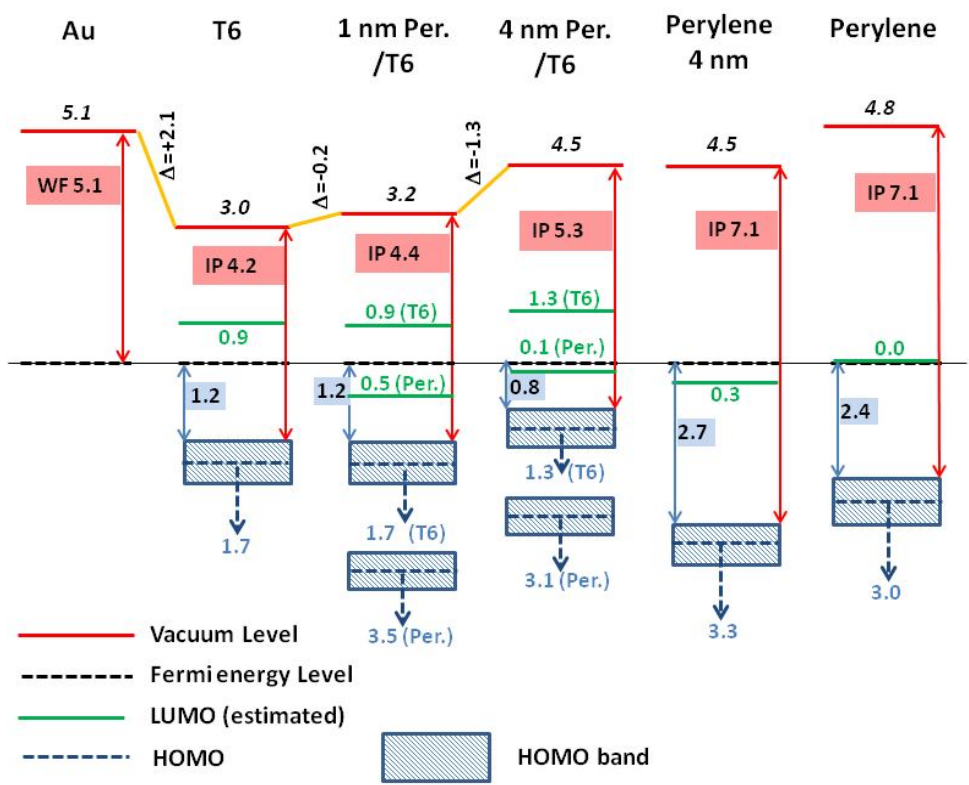

**FIG. 4** Energy level scheme of T6/PDI-8CN$_2$ heterojunctions, Au substrate, T6 and PDI-8CN$_2$ (4 nm and thick film). WF, organic film IP and the extrapolated LUMO position are also shown. All values are in eV. Energy level values are reported as distance from the Fermi Level. The HOMO band and centroid position are shown.